\documentclass[twocolumn,showpacs,preprintnumbers,amsmath,amssymb]{revtex4}

\usepackage{graphicx}
\usepackage{dcolumn}
\usepackage{bm}

\begin{document}

%\preprint{UIA/11.26}

\title{Symmetric and asymmetric states
in a mesoscopic superconducting wire in the voltage driven regime}

\author{D.Y. Vodolazov$^{1,2}$}
\email{vodolazov@ipm.sci-nnov.ru}
\author{F.M. Peeters$^2$}
\email{francois.peeters@ua.ac.be} \affiliation{$^1$ Institute for
Physics of Microstructures, Russian Academy of Sciences, 603950,
Nizhny Novgorod, GSP-105, Russia \\
$^2$Departement Fysica, Universiteit Antwerpen (CGB),
Groenenborgerlaan 171, B-2020 Antwerpen, Belgium}

\date{\today}

\pacs{74.25.Op, 74.20.De, 73.23.-b}

\begin{abstract}

The response of a mesoscopic homogeneous superconducting wire,
connected with bulk normal metal reservoirs, is investigated
theoretically as function of the applied voltage. The finite
relaxation length of the nonequilibrium quasiparticle distribution
function $\overline{L_E}$ is included where we assumed that our
wire is in the dirty limit. We found that {\it both} symmetric and
asymmetric states can exist which are characterized by a
stationary symmetric and asymmetric distribution of the order
parameter with respect to the center of the wire. Current voltage
characteristics of the wire with length $L>\overline{L_E}$ being
in the symmetric state show a pronounced S-behavior. The
asymmetric state may exist only for voltages larger than some
critical value and coexist with the symmetric state in a finite
voltage interval. For wires with $L \sim \overline{L_E}$ the
asymmetric state survives up to higher values of the voltage than
the symmetric one and may exist both in the voltage and the
current driven regimes. We propose an experiment to observe
reversible switching between those stationary symmetric and
asymmetric states.

\end{abstract}

\maketitle

\section{Introduction}

Modifications of the quasiparticle distribution function $f(E)$ in
a superconductor will drastically influence its properties.
Injection(extraction) of the quasiparticles via a tunnel junction
suppresses(enhances) the superconducting gap \cite{Chi,Blamire}
and creates a charge imbalance in the sample \cite{Clarke};
external electro-magnetic radiation excites quasiparticles to
higher energy levels and can considerably increase the
superconducting gap and the critical current
\cite{Dmitriev,Eliashberg}; moving vortices will change $f(E)$
resulting in the dependence of the vortex viscosity on the vortex
velocity \cite{Larkin}; fast oscillations of the superconducting
gap in superconducting bridges or in phase slip centers leads to a
dynamical nonequilibrium distribution $f(E)$ and results in
strongly nonlinear current-voltage characteristics \cite{Tinkham}.

Recently, the effect of nonequilibrium $f(E)$ induced by an
applied voltage was studied in a dirty superconducting wire (whose
mean path length $\ell$ is smaller than the coherence length at
zero temperature $\xi_0$) that was connected to large normal metal
reservoirs with no tunnel barriers \cite{Keizer}. Theoretically it
was found that there is a critical voltage $V_c$ above which the
superconductor exhibits a first order transition to the normal
state. This effect is connected with a strong modification of the
quasiparticle distribution function $f(E)$ and especially its odd
part $f_L(E)=f(-E)-f(E)$, by the applied voltage. The applied
voltage leads to the excitation of quasiparticles to energy levels
near the energy gap resulting in a strong modification (i.e.
diminishing) of the order parameter.

In the present work we study the same system as Ref. \cite{Keizer}
but {\it include} the finite relaxation time $\tau_E$ of the
nonequilibrium quasiparticle distribution function $f(E)$ due to
inelastic electron-phonon interaction which was neglected in Ref.
\cite{Keizer}. We show, that the finite length
$\overline{L_E}=\sqrt{\tau_E D}$ ($D$ is the diffusion constant)
weakens the effect of the applied voltage on the superconducting
properties and changes qualitatively the shape of the IV
characteristic (it becomes S-shaped) already for wires with length
$L$ of several $\overline{L_E}$. Furthermore we find that in our
homogeneous wire there are stationary states with both symmetric
and asymmetric distributions of the order parameter (with respect
to the center of the wire). The asymmetric state exists only at
voltage larger than some critical value and can coexist with the
stationary symmetric states in a finite voltage interval. For
wires with length $L \sim \overline{L_E}$ it survives up to larger
applied voltages than the symmetric one and at specific conditions
even for larger currents.

The paper is organized as follows. In section II we present our
theoretical model. In sections III and IV we discuss the existence
of the symmetric and asymmetric states, respectively. Finally, in
section V we present our conclusions and propose an experimental
setup to observe the reversible transitions between the predicted
symmetric and asymmetric states.

\section{Model}

We used a quasi-classical approach to calculate the nonequilibrium
properties of a superconductor in the dirty limit and restrict
ourselves to temperatures near $T_c$. This allows us to use the
Usadel equation \cite{Usadel} for the normal $\alpha (E)=\cos
\Theta$ and anomalous $\beta (E)=\sin \Theta$ quasi-classical
Green functions in relatively simple form
\cite{Schmid,Kramer,Watts-Tobin}
\begin{equation}%1
\frac{d^2\Theta}{dx^2}+((2iE-\frac{1}{L_E^2})-(\nabla\varphi)^2
\cos\Theta)\sin\Theta+2|\Delta|\cos\Theta=0.
\end{equation}
In the same limit the diffusive type equation for the space
dependence of the transverse (even in energy)
$f_T(E)=1-f(E)-f(-E)$ and longitudinal (odd in energy)
$f_L(E)=f(-E)-f(E)$ parts of the quasiparticle distribution
function $2f(E)=1-f_L(E)-f_T(E)$ are given by
\begin{subequations}
\begin{eqnarray}%2a
\nabla((N_1^2-R_2^2)\nabla f_L)+2N_2R_2\nabla \varphi\nabla f_T
\nonumber \\
-\frac{N_1}{L_E^2}(f_L-f_L^0)=0,
\end{eqnarray}
\begin{eqnarray}%2b
\nabla((N_1^2+N_2^2)\nabla f_T)+2N_2R_2\nabla \varphi\nabla f_L-
\nonumber \\
\frac{N_1}{L_E^2} (f_T-\phi\frac{\partial f_L^0}{\partial
E})-2N_2|\Delta|f_T=0.
\end{eqnarray}
\end{subequations}
Here $\varphi$ is the phase of the order parameter
$\Delta=|\Delta|e^{i \phi}$, $\phi$ is an electrostatic potential,
$N_1(E)+iR_1(E)=\cos \Theta(E)$, $N_2(E)+iR_2(E)=\sin\Theta(E)$
and $f_L^0(E)=\tanh(E/2T)$ is the odd part of the equilibrium
Fermi-Dirac distribution function of the quasiparticles. The
dimensionless length $L_E=\sqrt{D\tau_E}/\xi_0=\sqrt{\tau_E
\Delta_0/\hbar}$ defines the range over which the nonequilibrium
distribution of the quasiparticles may exist in the sample.

Within the same approach we have the rather simple self-consistent
stationary equation for the order parameter $\Delta$
\begin{equation}%3
a_1\frac{d^2\Delta}{dx^2}+(1-a_2|\Delta|^2+\Psi_1+i\Psi_2)\Delta=0,
\end{equation}
which is analogous to the Ginzburg-Landau equation \cite{Thinkham}
but with the additional terms
$\Psi_1=\int_0^{\infty}R_2(f_L-f_L^0)dE/|\Delta|$ and
$\Psi_2=\int_0^{\infty}N_2f_T dE/|\Delta|$. Because we are
interested in a stationary solution we removed all time-dependent
terms in Eq. (2-3) which are present in the original equations.
\cite{Schmid,Kramer,Watts-Tobin}

In Eqs. (1-3) the order parameter $\Delta$ is scaled by the
zero-temperature value of the order parameter $\Delta_0 \simeq
1.76 k_BT_c$ (in weak coupling limit), distance is in units of the
zero temperature coherence length $\xi_0=\sqrt{\hbar D/\Delta_0}$
and temperature in units of the critical temperature $T_c$.
Because of this choice of scaling the numerical coefficients in
Eq. (3) are $a_1 \simeq 0.69$ and $a_2 \simeq 0.33$. The current
is scaled in units of $j_0=\Delta_0/(\xi_0\rho_n e)$ and the
electrostatic potential is in units of $\phi_0=\Delta_0/e$
($\rho_n$ is the normal state resistivity and $e$ is the electric
charge).

The deviation of $f_L(E)$ from its equilibrium value $f_L^0$ may
considerably influence the value of the order parameter through
the term $\Psi_1$. The nonzero $f_T(E)$ is mainly responsible for
the appearance of the charge imbalance in the superconductor and
the conversion of the normal current to the superconducting one
and vice versus. The latter one occurs due to the Andreev
reflection process (term $2N_2|\Delta|f_T$ in Eq. (2b)) or/and due
to inelastic electron-phonon interaction (term $N_1(f_T-\phi
\partial f^0/\partial E)/L_E^2$ in Eq. (2b)).

The system of Eqs. (1-3) was numerically solved using an iterative
procedure. First Eq. (3) was solved by the Euler method (we add a
term $\partial \Delta/\partial t$ in the right hand side of Eq.
(3)) until $|\Delta|$ became time independent. Than Eqs. (1-2)
were solved from which we obtained new potentials $\Psi_1$ and
$\Psi_2$ which were then inserted in Eq. (3). This iterations
process was continued until convergence was reached. In some
voltage interval no convergence was reached and we identify it as
the absence of stationary solutions to Eqs. (1-3). Thus our
numerical procedure automatically checks the stability of the
found stationary symmetric and asymmetric states.

The numerical solution of Eqs. (1-3) converges much faster than
the equivalent but more complicated Eqs. (5,7,8,9) in Ref.
\cite{Keizer}. This allowed us to consider on the one hand a
rather large interval of parameters but on the other hand it
restricts the validity of the obtained results to the temperature
interval $0.9<T/T_c<1$ which roughly corresponds to the validity
condition $\Delta(T)/k_BT_c \ll 1$ of Eqs.
(1-3)\cite{Watts-Tobin}.

Knowing the solution of Eqs. (1-3) the current in the system can
be found using the following equation
\begin{eqnarray} %4
j=2a_1|\Delta|^2\nabla \varphi
\nonumber \\
+\int_0^{\infty}\left( (N_1^2+N_2^2)\nabla f_T +2N_2R_2f_L\nabla
\varphi \right) dE.
\end{eqnarray}
The first term in Eq. (4) may be identified as a superconducting
current and the second one as a normal one. Assuming that the
effect of the free charges is negligible in the superconductors
the electrostatic potential is determined by the following
expression
\begin{equation} %5
\phi=\int_0^{\infty}N_1f_TdE.
\end{equation}
We used the following boundary conditions for the system of Eqs.
(1-3)
\begin{widetext}
\begin{subequations}%6
\begin{eqnarray}
\Theta(\pm L/2)=\Delta(\pm L/2)=0,
 \\
f_{L}(\pm L/2)=\frac{1}{2}\left(\tanh\left(\frac{E+V(\pm
L/2)}{2T}\right) + \tanh\left(\frac{E-V(\pm
L/2)}{2T}\right)\right),
 \\
f_{T}(\pm L/2)=\frac{1}{2}\left(\tanh\left(\frac{E+V(\pm
L/2)}{2T}\right) - \tanh\left(\frac{E-V(\pm
L/2)}{2T}\right)\right),
\end{eqnarray}
\end{subequations}
\end{widetext}
which models the situation where there is a direct
electrical contact of our superconducting wire of length L with
large normal metal reservoirs at an applied voltage $V(\pm L/2 )$.

\section{Symmetric states}

In Fig. 1 we present our calculated current-voltage (IV)
characteristics of the superconducting wire with length $L=100$,
temperature $T=0.98$ and different $L_E$. In this section we
consider the case of symmetric applied voltage $V(\pm L/2)=\mp V$.
\begin{figure}[hbtp]
\includegraphics[width=0.48\textwidth]{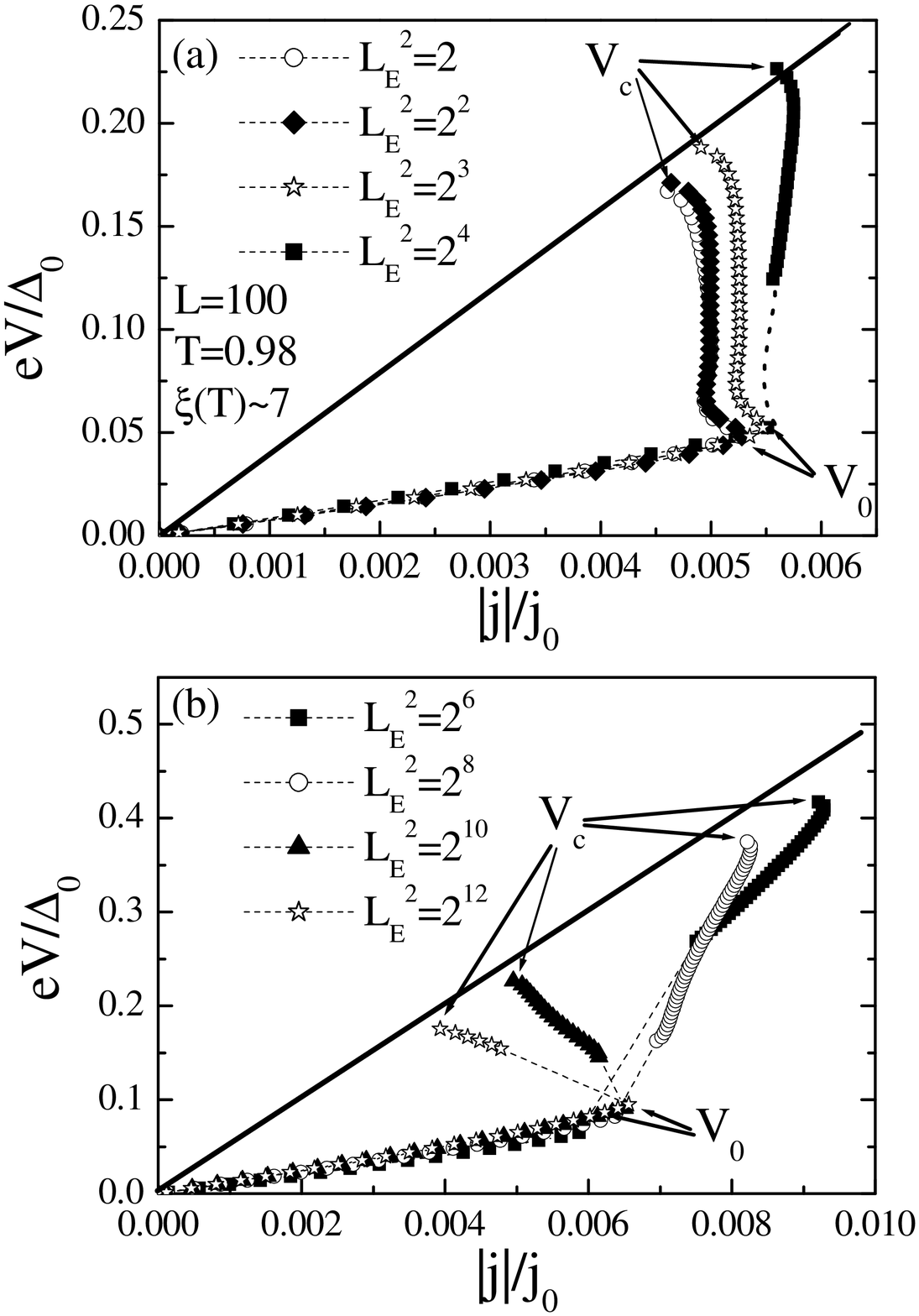}
\caption{\label{fig1} Current-voltage characteristics of the
superconducting wire with length $L=100$, temperature $T=0.98$ and
different relaxation lengths $L_E$. Figure (a) corresponds to the
case $L_E < \xi(T) \sim 7$ and figure (b) to the case $L_E \gg
\xi(T)$. Solid line shows the IV characteristic in the normal
state. Dashed lines in both (a) and (b) mark the voltage interval
where there are no stationary solutions to Eqs. (1-3).}
\end{figure}

When $L_E \lesssim \xi(T) \ll L$ the $f_L(E)$ decays on the scale
of the coherence length and practically does not influence the
value of the order parameter. When the current density in the wire
exceeds the critical value (marked by $V_0$ in Fig. 1), the N-S
borders (formed near the ends of the wire due to the boundary
condition Eq. (6c)) become unstable and start to move to the
center of the wire. The superconducting region decreases and it
effectively increases the resistance of the whole sample leading
to a decrease of the current up to its critical value. This
process results in the "vertical line"-like region in the IV
characteristics of the wire. The longer the sample, the wider will
be this "vertical line"-like region. Finally, at the critical
voltage $V_c$ the region where $|\Delta| \neq 0$ is reduced to
about $\xi(T)$ and a first order phase transition to the normal
state occurs.

In this limit we may use the so called 'local equilibrium model'
which leads to the extended time-dependent Ginzburg-Landau (TDGL)
equations \cite{Kramer,Watts-Tobin}. A simple analysis based on
those equations \cite{Vodolazov} shows an increasing critical
current with a stable N-S border when increasing the parameter
$\gamma=2\tau_E \Delta_0/\hbar = 2L_E^2$. This is the reason for
the increase of the maximal possible current in Fig. 1(a) for
small values of $L_E$. A direct comparison with results calculated
within the extended TDGL equations (see Fig. 2) show qualitatively
the same behavior.
\begin{figure}[hbtp]
\includegraphics[width=0.45\textwidth]{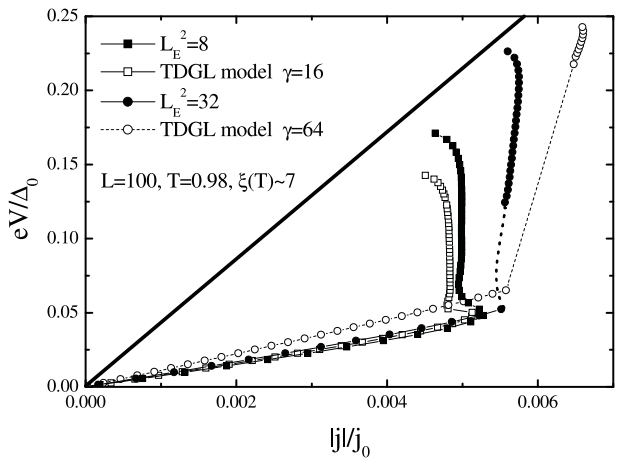}
\caption{\label{fig2} Comparison of the results obtained through
solution of Eqs. (1-3) with results of a numerical solution of the
extended TDGL equations. Up to $L_E\sim \xi(T)$ the extended TDGL
equations gives qualitatively the same results. Dashed lines mark
the voltage interval where no stationary solutions of Eqs. (1-3)
and the TDGL equations can be found.}
\end{figure}
\begin{figure}[hbtp]
\includegraphics[width=0.45\textwidth]{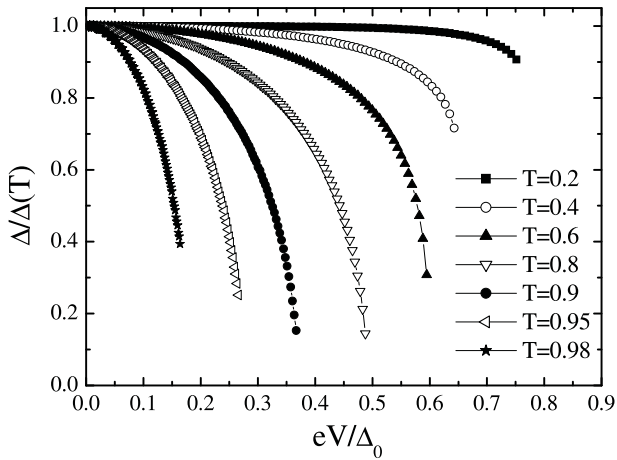}
\caption{\label{fig3} Dependence of the order parameter on the
parameter V as obtained from Eq. (7) for different temperatures.}
\end{figure}

In the opposite limit $L_E \sim L \gg \xi(T)$ the nonzero $f_L(E)$
considerably affects the value of the order parameter everywhere
in the sample. In Ref. \cite{Keizer} it was shown for $T=0$ and
$L_E=\infty$, that when the applied voltage is of about $\Delta_0$
the superconducting order parameter vanishes everywhere in the
sample and it leads to a first order phase transition to the
normal state. Actually, qualitatively the same is true for
arbitrary temperature: when the voltage reaches approximately
$\Delta(T)$ the order parameter starts to decrease rapidly. In
Fig. 3 we show the dependence of $|\Delta|(V)$ \cite{ours1} for a
spatially homogeneous distribution of $\Delta$ and $f_L(E)$ as
found from a numerical solution of the model equation for the
order parameter
\begin{equation} %7
1=\delta
\int_{|\Delta|}^{1/\sinh(\delta)}\frac{\tanh\left(\frac{E+V}{2T}\right)+\tanh
\left(\frac{E-V}{2T}\right)} {2\sqrt{E^2-|\Delta|^2}}dE,
\end{equation}
with $\delta=N(0)\nu/\Delta_0=0.3$ ($N(0)$ is the density of
states of electrons at Fermi surface and $\nu$ is a coupling
constant) and where instead of the equilibrium weight factor due
to the quasiparticles $\tanh(E/2T)$ \cite{Thinkham}, we use the
nonequilibrium one for the case of half voltage drop $V$.

At low temperatures the order parameter practically does not vary
with $V$ up to some critical voltage and it results in the absence
of that parts in the IV characteristics with negative differential
resistance  (see Ref. \cite{Keizer}). Contrary, at higher
temperatures, $|\Delta|$ may vary strongly with $V$ (see Fig. 3)
and it leads to a decay of the full current in the sample because
effectively the resistance of the wire grows with increasing the
voltage. It results in the appearance of a pronounced part with
negative differential resistance dV/dI in the IV characteristic
(see Fig. 2 in Ref. \cite{Keizer} and Fig. 1(b) for $L_E^2=2^{12}$
and $L_E^2=2^{10}$) for $V>V_0$. No stationary solutions to Eq.
(1-3) were found for samples with length larger than approximately
$10 \xi(T)$ at $V\gtrsim V_0$ and temperatures not far from $T_c$
(see illustration in Fig. 4 for $T=0.98$). This is connected with
the fact that the superconducting current reaches a value close to
the depairing current density in long samples but for shorter ones
the negative second derivative $d^2|\Delta|/d^2x$ starts to play
an essential role in Eq. (3) and stabilizes the stationary state
with nonzero $|\Delta|$.
\begin{figure}[hbtp]
\includegraphics[width=0.45\textwidth]{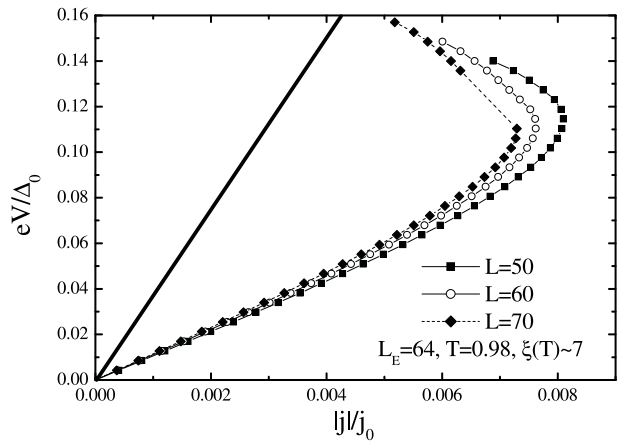}
\caption{\label{fig4} Current-voltage characteristics of the
superconducting wire with relaxation length $L_E=64$, temperature
$T=0.98$ and different wire lengths L (dashed line marks the
voltage interval with no stationary solutions of Eqs. (1-3)). It
demonstrates the absence of the stationary states at $V\sim V_0$
with increasing length of the wire.}
\end{figure}

For intermediate wire lengths $L \gg L_E \gg \xi(T)$ the situation
is more complicated. At $V=V_0$ due to the suppression of the
$|\Delta|$ near the edges by $f_L(E)$ the order parameter decays
and the current decreases with increasing voltage. In Fig. 5 we
show the spatial distribution of the order parameter for different
values of the voltage for $L_E=64$ and $L_E=16$ (L=100, T=0.98).
Due to the decay of $f_L$ on the scale $L_E$ and the term $\Psi_1$
in Eq. (3), the order parameter also starts to vary on a distance
of about $L_E$. That's why the order parameter is finite in a
wider region of the wire with $L \sim L_E$ than for wires with
$L\gg L_E$ even at $V \sim V_c$. The suppression of $|\Delta|$
near the ends of the short wire (with respect to $L_E$) is mainly
connected with the relaxation of $f_L(E)$ due to the coupling with
the transverse mode in Eq. (2a) (see discussion below).
\begin{figure}[hbtp]
\includegraphics[width=0.45\textwidth]{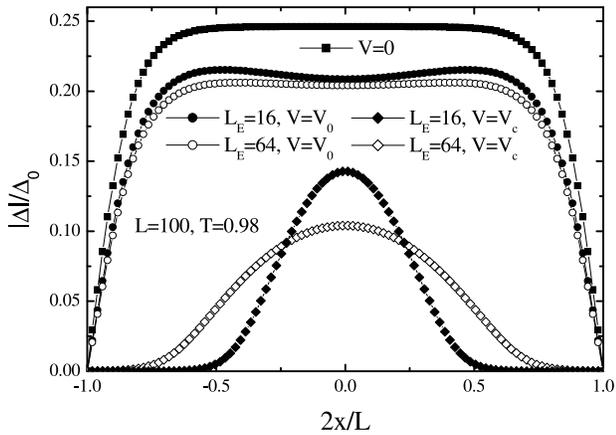}
\caption{\label{fig5} Distribution of the order parameter in a
superconducting wire of length $L=100$ being in the symmetric
superconducting state at different values of the applied voltage
and relaxation lengths $L_E$.}
\end{figure}
\begin{figure}[hbtp]
\includegraphics[width=0.45\textwidth]{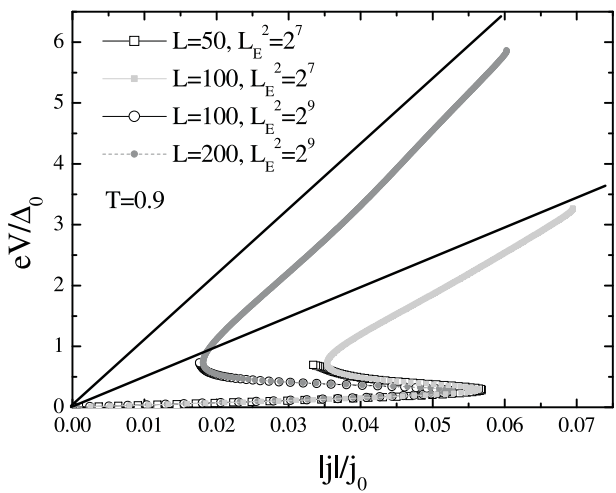}
\caption{\label{fig6}Current-voltage characteristics of the
superconducting wires of different lengths and electron-phonon
relaxation time (solid curves correspond to the normal state of
the wires with lengths $L=100$ and $L=200$). It shows, that for a
fixed temperature (not too close to $T_c$ in order to satisfy the
condition $L_E \gg \xi(T)$) and relaxation length $L_E$ it is
possible to find the length of the sample for which the IV
characteristic will have a pronounced S-behavior.}
\end{figure}
When the region of the suppression of $|\Delta|$ increases up to
about $L_E$ the effect of the nonequlibrium $f_L(E)$ becomes less
pronounced, $|\Delta|$ decreases much slower with increasing
voltage and the current starts to increase again. At some moment
it reaches a value close to the depairing current density in the
center of the wire and the state with a nonequlibrium and a
stationary distribution of the order parameter becomes unstable
\cite{ours2}. Phase slip centers should appear in the wire. This
is observed as the absence of a stationary solution to Eqs. (1-3)
in some range of applied voltage (marked dashed lines in Fig.
1(b)). With further increasing voltage the superconducting region
decreases and we find again a stationary solution. At this
voltage, the size of the superconducting region is too small to
'carry' a phase slip state and is close to the critical size $10
\xi(T)$ found above for the stability of the stationary state in
the wire with $L_E \sim L$. The above mechanism leads to a S-shape
for the IV characteristics for the intermediate case $L \gg L_E
\gg \xi(T)$. This feature is more visible if we consider low
temperatures and study samples of different lengths (see Fig. 6).

The relaxation of $f_L(E)$ and $f_T(E)$ to equilibrium occurs not
only via the inelastic electron-phonon interaction but also due to
the coupling terms $2DN_2R_2\nabla \varphi \nabla f_{T,L}$ in Eqs.
(2a,b). They are nonzero in the region where the order parameter
changes near the N-S border and where $\nabla f_{T,L} \neq 0$.
This mechanism effectively relaxes the nonequilibrium $f_{L,T}(E)$
at energies close to $|\Delta|$ (where the product $N_2R_2 \neq
0$). In Fig. 7 we present the spatial dependence of
$f_L(E,x)-f_L^0(E)$ and $f_T(E,x)$ for different values of the
energy for a wire with length L=100, temperature T=0.98,
relaxation length $L_E=64$, applied voltage $V=0.08$ and with zero
and nonzero coupling terms $2DN_2R_2 \nabla \varphi \nabla
f_{T,L}$ in Eq. (2). We notice that for energies close to
$|\Delta|$ the coupling of the longitudinal mode with the
transverse one leads even to a change in sign of $f_L(E)$ in the
central part of the sample and hence to a smaller value of the
term $\Psi_1$ in Eq. (3). The change of the relaxation length of
the transverse mode $f_T$ is not so pronounced, but is still
noticeable and justifies the faster decay due to coupling with the
longitudinal mode.
\begin{figure}[hbtp]
\includegraphics[width=0.48\textwidth]{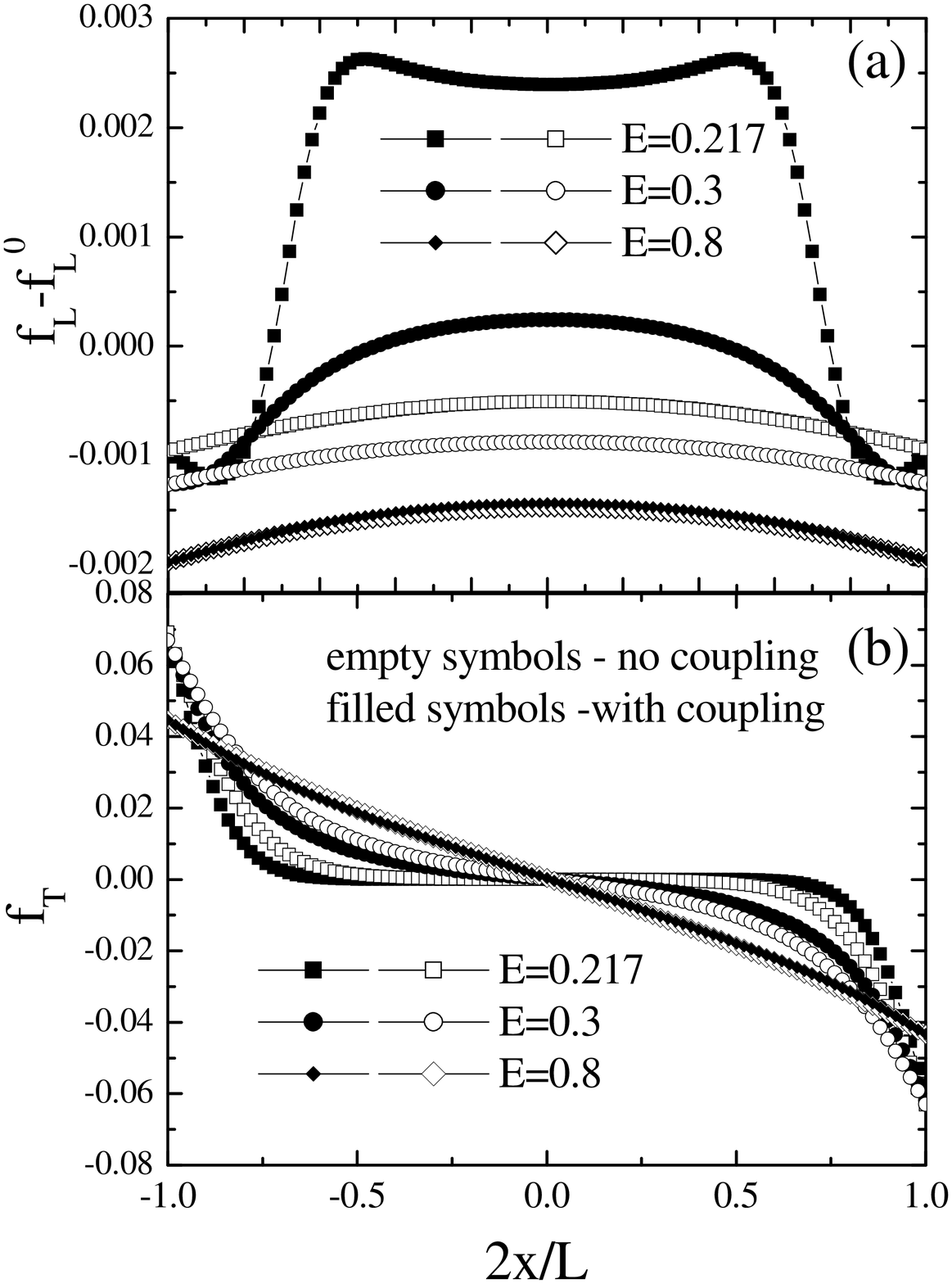}
\caption{\label{fig7} Spatial dependence of $f_L-f_L^0$ (a) and
$f_T$ (b) at different energies for zero and nonzero coupling
terms $2N_2R_2 \nabla \varphi \nabla f_{L,T}$ in Eqs. (2a,b).
Parameters of the wire: L=100, $L_E^2=2^9$, $T=0.98$, $V=0.08$,
$\Delta(x=0,V=0.08)\simeq 0.216$.}
\end{figure}

That effect provides the dependence of the shape of the
current-voltage characteristic of the wire on temperature (see
Fig. 8) for fixed length of the sample and strength of the
electron-phonon interaction even when $L \gg L_E >\xi(T)$. Above
discussed additional mechanism of relaxation effectively shortens
the relaxation length of $f_{L,T}(E)$ for $V \gtrsim V_0 $ and
decreases the effect of nonequilibrium when $L_E$ approaches
$\xi(T)$. It moves the wire from the limit $L \gg L_E \gg \xi(T)$
to the limit $L \gg L_E \sim \xi(T)$ at some temperature close to
$T_c$.
\begin{figure}[hbtp]
\includegraphics[width=0.48\textwidth]{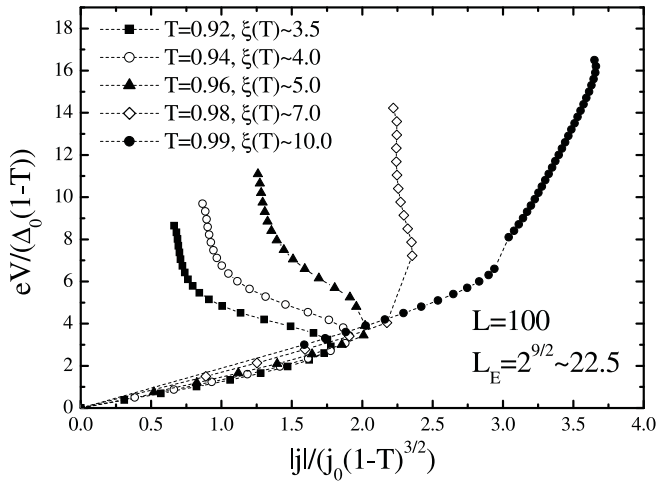}
\caption{\label{fig8} Current-voltage characteristics of the
superconducting wire of fixed length and strength of the
electron-phonon interaction for different temperatures. It shows
the change of the shape of the IV characteristic when $L_E$
approaches $\xi(T)$.}
\end{figure}

\section{Asymmetric states}

To understand the origin of the asymmetric states in a
superconducting wire let us consider as a example the mesoscopic
normal metal wire half of which has a resistance larger than the
other half. Applying a voltage to such a sample the voltage drop
in the high resistance part will be larger than in the low
resistance part (for definition we suppose that $R_{left}
> R_{right}$) in order to satisfy the continuity of the current along
the wire, i.e. $I(x)={\rm const}$. It means that if we put V=0 in
the center of our inhomogeneous wire we have $|V(-L/2)|>|V(L/2)|$.
Hence the deviation of the quasiparticle distribution functions
from equilibrium in the left and right normal reservoirs ($f(E,x
\pm L/2)-1/(\exp(E/T)+1)$) will depend not only on the value of
the applied voltage but also on the properties of the mesoscopic
wire.
\begin{figure}[hbtp]
\includegraphics[width=0.48\textwidth]{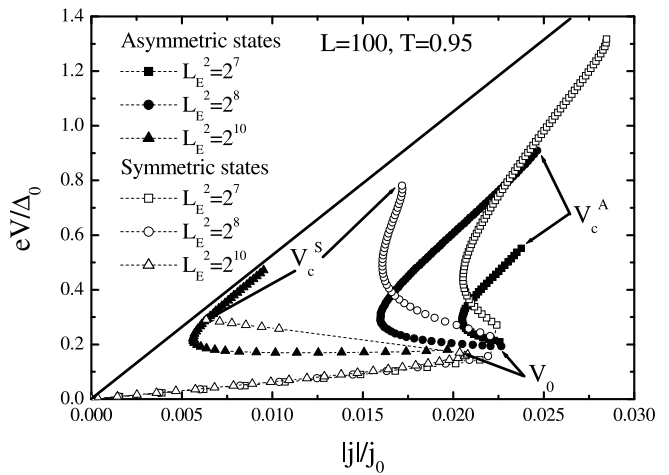}
\caption{\label{fig9} Current-voltage characteristics of the
symmetric (empty symbols) and asymmetric (filled symbols) states.
Solid curve shows the normal state. Dashed curves mark regions of
absence of the stationary symmetric states. The stationary
asymmetric state does not exist for $V<V_0$ - voltage where
negative-differential resistance appears for symmetric state.}
\end{figure}

To find corresponding superconducting state in our {\it
homogeneous} system we use the asymmetric boundary conditions
$V(\pm L/2)=\mp V+dV$. The addition dV is determined from the
condition of constant current along the wire. It increases from
zero (at $V \simeq V_0$) to a finite value (at $V_c^{a}$ -
critical voltage of the asymmetric state). No asymmetric states
were found for $V<V_0$. In Fig. 9 we show typical IV
characteristics and in Fig. 10 we present the corresponding
distribution of the order parameter for symmetric and asymmetric
states at the same values of the voltage. It is obvious from the
above discussion that the order parameter is more suppressed on
the side of the wire with maximal absolute value of the voltage.
For example if we change the sign of dV the order parameter
distribution in Fig. 10 will be symmetrically reflected with
respect to x=0.

It is interesting to note that such an asymmetric state may exist
up to larger values of the voltage when $L \gtrsim L_E \gg \xi$,
and even for larger values of the current, than the symmetric
state (latter property is realized for lengths $L \sim L_E$ for
which the IV characteristic changes slope from a negative to a
positive one for the symmetric state when $V>V_0$). Another
interesting property is the existence of the stationary asymmetric
state for voltages when the stationary symmetric state is absent
(see Fig. 9).

\begin{figure}[hbtp]
\includegraphics[width=0.48\textwidth]{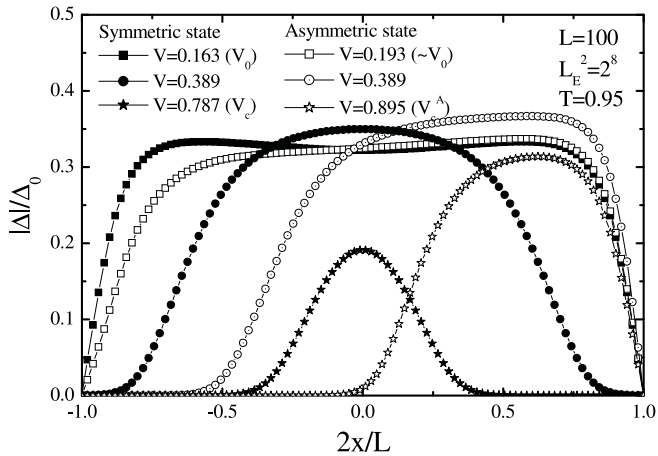}
\caption{\label{fig10} Order parameter distribution for asymmetric
and symmetric states.}
\end{figure}

The range of the existence of the stationary asymmetric state is
rather small in comparison with the symmetric state when the
length of the wire is much larger than $L_E$. It exists only at
voltages near $V_0$ (see Figs. 9 and 11 for small ratios $L_E/L$).

We did not find stationary asymmetric states for wires with $L_E
\sim \xi(T)$. For small values $L_E \ll \xi(T)$ the asymmetric
state is nearly the same as the symmetric one but shifted with
respect to the point $x=0$ and the IV characteristics are found to
be nearly the same for both symmetric and asymmetric states. This
is explained by the negligible effect of $f_L(E)$ on the value of
the order parameter in this limit.
\begin{figure}[hbtp]
\includegraphics[width=0.48\textwidth]{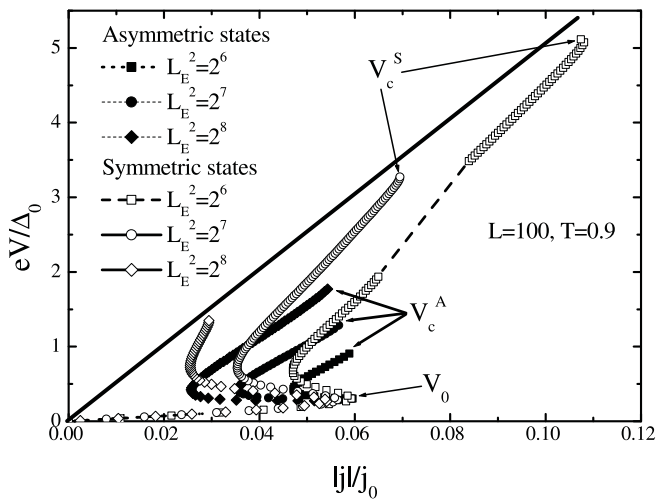}
\caption{\label{fig11} Current-voltage characteristics for
symmetric and asymmetric states for wires with different $L_E$.}
\end{figure}

\section{Conclusion}

The response of a dirty superconducting wire that is connected to
normal metal banks, to an applied voltage, strongly depends on the
ratio between the length of the sample $L$, the coherence length
$\xi$ and relaxation length $\overline{L_E}$ of the nonequilibrium
quasiparticle distribution function $f(E)$. The latter one, was
determined in our calculations by the inelastic electron-phonon
interaction $\overline{L_E}=\sqrt{D\tau_E}$ with an energy
independent characteristic time $\tau_E$. Applied voltage
suppresses the superconducting order parameter both via the
creation of the superconducting current and the modification of
the quasiparticle distribution function. First mechanism is
prevalent when $L_E \lesssim \xi(T)$ and the second one is
dominant in the opposite limit $L_E \gg \xi(T)$ for samples with
$L \sim L_E$. For wires with $L \gg L_E \gg \xi(T)$ the transition
to the normal state is mainly induced by the superconducting
current.

We found that for the following two cases $L \gg \xi(T)
> L_E$ and $L \gtrsim L_E \gg \xi(T)$ there are three stationary
solutions to Eqs. (1-3) at given value of the applied voltage
difference $2V>2V_0$ - one symmetric state and two asymmetric
states (which are symmetrically reflected to the center of the
wire) which are characterized by symmetric and asymmetric
distribution of the order parameter, respectively. The degeneracy
is most pronounced in the limit $L \gtrsim L_E \gg \xi(T)$ because
the order parameter may vary on the distance of order $L_E \gg
\xi$ due to the long relaxation length of the odd (in energy) part
of the nonequilibrium quasiparticle distribution function and it
provides the basis for the appearance of new effects. For example
it leads to an S-behavior of the IV characteristic of the wire
being in the symmetric state. In the same limit the stationary
asymmetric state may exist even when the symmetric one exist only
as a time-dependent one (phase slip state) (see Fig. 9). Moreover,
for a specific temperature and length $L_E \gg \xi(T)$ there exist
length of the sample for which a stationary asymmetric state does
exist up to a larger value of the current than the stationary
symmetric one (see Fig. 9 and compare for $L_E^2=2^8$).

We should mention here, that in the extended TDGL equation there
is no direct effect of the voltage on the value of the order
parameter. This is because of the "local equilibrium
approximation" (valid in the limit $L_E \ll \xi (T)$)
\cite{Kramer,Watts-Tobin} where it was assumed that $f_L-f_L^0$ is
only proportional to the time variation of the value of the order
parameter $\partial |\Delta|/\partial t$ and hence is zero in the
stationary case. But it is interesting to note that the asymmetric
state could also be obtained by this equation. For example the
asymmetric state exists for $\gamma=32, L=100, T=0.98$ (see inset
in Fig. 1a) at $0.217<V<0.242$. But we did not find in this model
the wire parameters for which asymmetric and symmetric states
could exist at the same value of the voltage.
\begin{figure}[hbtp]
\includegraphics[width=0.48\textwidth]{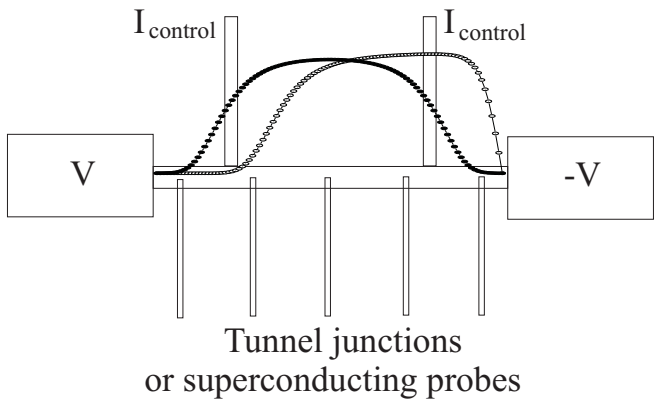}
\caption{\label{fig12} Proposed experimental setup to measure
transitions between symmetric and asymmetric states and
current-voltage characteristic of both states. The curves with
empty and filled symbols show the order parameter distribution for
a wire with parameters from Fig. 10 and V=0.389.}
\end{figure}

Experimentally the asymmetric state may be realized by adding
additional contacts to the superconducting wire with control
current close to the ends of the wire (see Fig. 10). It is better
to contact the wire with the reservoirs made from the same
material and to apply a strong enough magnetic field to suppress
the superconductivity in the reservoirs. This procedure will
provide good contact with normal reservoirs. Applying strong
enough current to only one of the current contacts we locally
destroy superconductivity and force an asymmetric distribution of
the order parameter. After switching off the control current the
asymmetric distribution should be stable (at proper choice of the
working point at the IV characteristic). To come back to the
symmetric state we should apply a control current at the both
current contacts and it will force the recovering the symmetric
state in the wire. We will need a series of tunnel junctions along
the wire to measure locally the strength of the order parameter.
Other alternative is to use superconducting (SC) probes to
distinguish the part of the superconductor with attached SC-probe
in the normal or superconducting state. When in the normal state
the superconducting probe measures the electrostatic potential.

Good candidates to observe these symmetric and asymmetric states
is dirty aluminium and zinc with their relatively large coherence
lengths ($\xi_{Al}(0) \sim 0.15 \mu m$, $\xi_{Zn}(0) \sim 0.25 \mu
m$ ) and $\overline{L_E}(Al) \sim 9 \mu m $, $\overline{L_E}(Zn)
\sim 35 \mu m $ ($L_E(Al) \sim 60 $, $L_E(Zn) \sim 140 $). The
expected temperature range for the validity of our results is
$0.9<T/T_c<1$. The most strict restriction to observe these
effects is that there should be only a small heating of the
sample. Heating will initiate the transition of the sample to the
normal state and hide the main effects.

For other low-temperature superconductors (Nb, Pb, In, Sn) we have
$L_E < 10$ (we used data for $\tau_E$ at $T\sim T_c$ from Ref.
\cite{Stuivinga}). It means that at $T \geq 0.9$ the coherence
length is comparable with $L_E$. For these conditions we did not
find asymmetric states. Moreover the effect of nonequilibrium in
the odd part of the quasiparticle distribution function that is
connected with the applied voltage is rather small for these
parameters and the extended TDGL equations gives result which are
qualitatively similar to the one obtained here.

\begin{acknowledgments}
This work was supported by the Flemish Science Foundation
(FWO-Vl), the Belgian Science Policy (IUAP) and the ESF-AQDJJ
program. D. Y. V. acknowledges support from INTAS Young Scientist
Fellowship (04-83-3139).
\end{acknowledgments}


\begin{references}

\bibitem{Chi} C. C. Chi and J. Clarke, Phys. Rev. B {\bf 20}, 4465 (1979).

\bibitem{Blamire} M. G. Blamire, E. C. G. Kirk, J.E. Evetts, and
T. M. Klapwijk, Phys. Rev. Lett. {\bf 66}, 220 (1991).

\bibitem{Clarke} J. Clarke, in: {\it Nonequlibrium superconductivity},
edited by D.N. Langenberg and A.I. Larkin, (Elsevier Science
Publisher B.V., Berlin, 1986), p. 1 and references therein.

\bibitem{Dmitriev} V. M. Dmitriev, V. N. Gubankov and F. Ya. Nad',
in: {\it Nonequlibrium superconductivity}, edited by D.N.
Langenberg and A.I. Larkin, (Elsevier Science Publisher B.V.,
Berlin, 1986), p. 163 and references therein.

\bibitem{Eliashberg} G. M. Eliashberg and B. I. Ivlev, in:
{\it Nonequlibrium superconductivity}, edited by D.N. Langenberg
and A.I. Larkin, (Elsevier Science Publisher B.V., Berlin, 1986),
p. 211 and references therein.

\bibitem{Larkin} A.I. Larkin and Yu. N. Ovchinnikov, in:
{\it Nonequlibrium superconductivity}, edited by D.N. Langenberg
and A.I. Larkin, (Elsevier Science Publisher B.V., Berlin, 1986),
p. 493 and references therein.

\bibitem{Tinkham} M. Tinkham, {\it Introduction to
superconductivity}, (McGraw-Hill, NY, 1996).
% , p. 417 and references therein.

\bibitem{Keizer} R. S. Keizer, M. G. Flokstra, J. Aarts, and T. M.
Klapwijk, Phys. Rev. Lett. {\bf 96}, 147002 (2006).

\bibitem{Usadel} K. D. Usadel, Phys. Rev. Lett. {\bf 25}, 507 (1970).

\bibitem{Schmid} A. Schmid and G. Sch\"on, J. Low Temp. Phys. {\bf
20}, 207 (1975).

\bibitem{Kramer} L. Kramer and R.J. Watts-Tobin, Phys. Rev. Lett.
{\bf 40}, 1041 (1978).

\bibitem{Watts-Tobin} R.J. Watts-Tobin, Y. Kr\"ahenb\"uhl, and L. Kramer,
 J. Low Temp. Phys. {\bf 42}, 459 (1981).

\bibitem{Vodolazov} D. Y. Vodolazov, A. Elmurodov, and F. M. Peeters,
Phys. Rev. B {\bf 72}, 134509 (2005).

\bibitem{ours1} We found that for $T<0.25$ there exist two solutions
to Eq. (7). Because we are not interested in this temperature
interval we did not investigate this phenomena in detail.

\bibitem{ours2} Here we should mention that for relatively
large relaxation length $L_E > \xi(T)$ the critical current for
the stability of the N-S border is larger than the depairing
current density and superconductivity becomes unstable in the
center of the wire. The reason is that the superconducting current
(not the whole current) destroys superconductivity, but it is
small at the N-S border for large $L_E$.

\bibitem{Stuivinga} M. Stuivinga, J. E. Mooij, and T. M. Klapwijk,
J. Low Temp. Phys. {\bf 46}, 555 (1982).

\end{references}
\end{document}